\documentclass[showpacs,twocolumn,eqsecnum,prb,amsmath,amssymb]{revtex4}
\usepackage{graphicx}% Include figure files
\usepackage{dcolumn}% Align table columns on decimal point
\usepackage{bm}% bold math
\begin{document}
\title{
Kondo-lattice theory of anisotropic singlet superconductivity
}
\author{Fusayoshi J. Ohkawa}
\affiliation{Department of Physics, Faculty of Science,  
Hokkaido University, Sapporo 060-0810, Japan}
\email{fohkawa@phys.sci.hokudai.ac.jp}
%\date{\today}
\received{January 24, 2008}
\begin{abstract} 
Kondo-lattice theory for the $t$-$J$ model and a phenomenological theory based on it are developed to study superconductivity in the vicinity of the Mott metal-insulator transition.
Since the quenching of magnetic moments by single-site quantum spin fluctuations or the Kondo effect is reduced by the opening of a superconducting gap, spin density wave (SDW) can appear in a superconducting state and the Knight shift can deviate from the Yosida function to be small.
The electron-phonon interaction, which arises from the modulation of the superexchange interaction by phonons, is crucial in the coexistence of superconductivity and SDW.
It is proposed that the coexistence of superconductivity, a double-${\bf Q}$ SDW, a double-${\bf Q}$ lattice distortion,  and a double-${\bf Q}$ charge density wave induced by the SDW rather than the lattice distortion, is responsible for the checkerboard structure and the zero-temperature pseudo-gap observed in under-doped cuprate superconductors.
\end{abstract}
 % 71.10.-w Theories and models of many electron systems
 % 71.10.Ay Fermi-liquid theory and other phenomelological models 
 % 71.27.+a Strongly correlated electron systems; heavy fermions
 % 71.30.+h Metal-insulator transitions 
 %          and other electronic transitions
 % 74.20.-z Theories and models of superconducting state
 % 74.90.n, Other topics in superconductors
 %          (restricted to new topics in section 74) 
 % 75.10.-b General theory and models of magnetic ordering
 % 75.10.Lp Band and itinerant models
 % 75.30.Et Exchange and superexchange interactions
 % 75.30.Kz Magnetic phase boundaries
 % 75.30.Gw Magnetic anisotropy
 % 75.50.-y Studies of specific magnetic materials
%%%%%%%%%%%%%%%%%%%%%%%%%%%%%%%%%%%%%%%%
\maketitle
\section{Introduction}
High temperature (high-$T_c$) superconductivity is one of the most interesting and important issues in condensed-matter physics. \cite{bednorz,RevD-wave,RevScience,RevStripe,lee,RevScanning}
It occurs in highly anisotropic quasi-two dimensional cuprate oxides, i.e., on CuO$_2$ planes. 
Parent cuprates are Mott insulators. 
Antiferromagnetism occurs at low temperatures because of the superexchange interaction between nearest neighbor Cu ions on CuO$_2$ planes, which is as strong as \cite{SuperJ}
\begin{equation}\label{EqExpJ}
J=- \mbox{(0.10--0.15)~eV}.
\end{equation}
When {\it holes} or electrons are doped, cuprates 
become metals and $d_{x^2-y^2}$-wave or 
$d\gamma$-wave superconductivity occurs on CuO$_2$ planes. 
High-$T_c$ superconductivity
occurs in the vicinity of the Mott metal-insulator (\mbox{M-I}) transition.

The Hubbard model is one of the simplest effective Hamiltonians
for the Mott \mbox{M-I} transition.
On the basis of Hubbard's \cite{Hubbard1,Hubbard2}
and Gutzwiller's \cite{Gutzwiller1,Gutzwiller2,Gutzwiller3} theories,
which are under the single-site approximation (SSA), it is speculated that
the density of state (DOS) has a three-peak structure, with
the Gutzwiller band between the upper Hubbard band (UHB)
and the lower Hubbard band (LHB).
Another SSA theory confirms this speculation, \cite{OhkawaSlave} showing
that the Gutzwiller band appears at the top of LHB for $n<1$, with
$n$ being the electron density per unit cell.
The supreme SSA, which considers all the single-site terms, is reduced
to determining and solving self-consistently the Anderson model, 
\cite{Mapping-1,Mapping-2,Mapping-3}
which is an effective Hamiltonian for the Kondo effect.
The three-peak structure corresponds to 
the Kondo peak between two subpeaks in the Anderson model.
The Kondo effect has relevance to the Mott \mbox{M-I} transition.

The supreme SSA is rigorous for infinite dimensions within the restricted Hilbert subspace where no order parameter exists. \cite{Metzner}
It is later shown that
the supreme SSA can also be formulated as the dynamical mean-field theory\cite{georges,RevMod,kotliar,PhyToday} (DMFT) and the dynamical coherent potential approximation \cite{dpca} (DCPA); 
they are exactly the same approximation as each other.  

As a function of doping concentrations $\delta=|1-n|$, superconducting (SC) critical temperatures $T_c$ show a maximum at an optimal 
one $\delta\simeq 0.15$, which is denoted by $\delta_{\rm op}$. 
Cuprates with $\delta \alt \delta_{\rm op}$,
$\delta \simeq \delta_{\rm op}$, and $\delta \agt \delta_{\rm op}$ are
called under-doped, optimal-doped, and over-doped ones, respectively.
Specific heat coefficients of optimal-doped ones
are as large as \cite{gamma1,gamma2,gamma3} 
\begin{equation}\label{EqObsGamma}
\gamma =10\mbox{--}
15~\mbox{mJ/K}^2\cdot \mbox{CuO}_2\mbox{-mol} ,
\end{equation}
which are consistent with the prediction of Gutzwiller's theory.
\cite{Gutzwiller1,Gutzwiller2,Gutzwiller3} 
From Eq.~(\ref{EqObsGamma}), 
the bandwidth of quasi-particles is estimated to be
\begin{equation}
W^* =0.3\mbox{--}0.5~\mbox{eV}.
\end{equation}
Cooper pairs can also be bound by a magnetic exchange interaction. 
\cite{hirsch}  
One of the simplest effective Hamiltonians for cuprates is
the $t$-$J$ model; \cite{RVB} it is derived from the Hubbard model \cite{hirsch} and is also derived from the the $d$-$p$ model,\cite{rice} which considers $3d$ orbits on Cu ions and $2p$ orbits on O ions.
It is quite easy to reproduce observed $T_c$
if the phenomenological $J$ and $W^*$ are used. 
It is proposed by an early theory \cite{highTc1,highTc2} that
the condensation of $d\gamma$-wave Cooper pairs between the Gutzwiller
quasi-particles is responsible for high-$T_c$ superconductivity.

%Anomalies are observed in cupartes.

There are pieces of evidence that the electron-phonon 
interaction is strong on CuO$_2$ planes:
the softening of the half-breathing modes
around $\left(\pm \pi/a, 0\right)$ and $\left(0, \pm \pi/a\right)$
in the two-dimensional Brillouin zone (2D-BZ),
\cite{McQ1,Pint1,McQ2,Pint2,Braden}
with $a$ the lattice constant of the CuO$_2$ planes,
the softening of Cu-O bond stretching modes
around $(\pm \pi/2a,0)$ and $(0,\pm \pi/2a)$ in 2D-BZ, 
\cite{pintschovius,reznik} 
kinks in the dispersion relation of quasi-particles, 
\cite{johnson,tsato} and so on.
On the other hand, observed isotope shifts of 
$T_c$ are small, \cite{isotope} which implies that
the strong electron-phonon interaction can play no significant role
in high-$T_c$ superconductivity itself. 

The modulation of the superexchange interaction by phonons gives a strong electron-phonon interaction. \cite{FJO-ph1,FJO-ph2} 
It can explain the softening of phonons. 
The virtual exchange of a relevant phonon gives a mutual attractive interaction. Since it is strong between next nearest neighbors
but is weak between nearest neighbors,
it can play no significant role in the binding of $d\gamma$-wave Cooper pairs, which are bound mainly between nearest neighbors.
 
Many anomalies are observed in under-doped cuprates; some of the most crucial ones are the pseudo-gap,\cite{RevScanning,yasuoka,Renner,Ido1,Ido2,Ido4,Ekino} the stripe or checkerboard structure,  \cite{RevStripe,RevScanning,howald,hoffman1,hoffman2,fang,Hanaguri,ido4*4} the zero-temperature pseudo-gap (ZTPG),\cite{RevScanning,Hanaguri,mcErloy1,mcErloy2} and so on.
One may argue that an exotic mechanism must be responsible for 
not only the anomalies but also superconductivity.
 Since $T_c$'s are more or less low when one or some of the anomalies are observed, the other may argue that the role of what cause the anomalies is simply the suppression of $T_c$. 
The anomalies should be consistently explained with high-$T_c$ superconductivity.

The supreme SSA is mapped to solving the Anderson model, as is discussed above. When no SC gap opens in the Hubbard, $t$-$J$, or $d$-$p$ model, the Fermi surface (FS) exists in the Anderson model so that the ground state of the Anderson model is the normal Fermi liquid (FL). In the SSA, the ground state is stabilized by the Kondo effect and is the FL. 
A perturbative treatment of intersite effects starting from the {\it non-perturbative} SSA theory is simply Kondo-lattice theory.
Beyond the SSA, the FL is further stabilized  by the Fock-type term of the superexchange interaction, \cite{FJO-Hartree,FJO-RVB} which favors a local singlet or a resonating valence bond (RVB) on each pair of nearest neighbors;  the stabilization mechanism is physically the same as the RVB mechanism. \cite{RVB,fezekas,Plain-vanilla}

Critical fluctuations make $T_c$ down to $+0$~K in two dimensions, \cite{mermin} which implies that they 
are also crucial in highly anisotropic quasi-two dimensions.
If the anisotropy is large enough but $T_c$ is still high enough, SC critical fluctuations can cause the opening of a pseudo-gap in the SC  critical region. \cite{FJO-PsGap1,FJO-PsGap2}

The Kondo effect is the quenching effect of magnetic moments by single-site quantum spin fluctuations, whose energy scale is the Kondo temperature $k_BT_{\rm K}$ or $T_{\rm K}$.
When an SC gap opens, 
the FS vanishes in the Anderson model 
so that $T_{\rm K}$ is reduced.
It is possible that  $T_{\rm K}$ is so reduced in an SC state that spin density wave (SDW) can appear in the SC state.
One of the purposes of this paper is to develop microscopic and phenomenological theories to study possible coexistence of superconductivity and SDW in under-doped cuprate superconductors.
The organization of this paper is as follows:
In Sec.~\ref{SecKL-theory}, Kondo-lattice theory of superconductivity is extended such that it can treat properly the reduction of $T_{\rm K}$ in an anisotropic SC state.
It is shown in Sec.~\ref{SecReductionTK} that, because of the reduction of $T_{\rm K}$, SDW can appear in the SC state and the Knight shift can deviate from the Yosida function.
In Sec.~\ref{SecSDW-lattice}, a phenomenological theory is developed on the basis of the microscopic theory developed in Sec.~\ref{SecKL-theory} to study the coexistence of superconductivity and SDW; it is shown that the electron-phonon interaction is crucial in the coexistence.
It is proposed in Sec.~\ref{SecAnomalies} that the coexistence is responsible for the checkerboard  and ZTPG observed in under-doped cuprate superconductors.
Discussion is given in Sec.~\ref{SecDiscussion}.
Conclusion is given in Sec.~\ref{SecConclusion}.
Anderson's compensation theorem is studied in
Appendix~\ref{AppCompensation}.

%\noindent 1 

\section{Kondo-lattice theory}
\label{SecKL-theory}
\subsection{Nambu representation}

In this paper, the $t$-$J$ or $t$-$J$-$U_\infty$ model 
on the square lattice is considered:
\begin{eqnarray}\label{Eqt-J}
{\cal H} &=&
\epsilon_a \sum_{i\sigma} n_{i\sigma}
+ \sum_{i \ne j\hskip2pt \sigma} t_{ij} a_{i\sigma}^\dag a_{j\sigma} 
- \frac1{2} J \sum_{\left<ij\right>} 
\left({\bf S}_i\cdot{\bf S}_j\right)
\nonumber \\ && \qquad
+ U_\infty \sum_{i} n_{i\uparrow}n_{i\downarrow}, 
\end{eqnarray}
with $n_{i\sigma}=a_{i\sigma}^\dag a_{i\sigma}$, $\epsilon_a$ the band center,  and 
\begin{equation}
{\bf S}_i = \sum_{\alpha\beta}
\frac1{2} 
\left(\sigma_x^{\alpha\beta}, \sigma_y^{\alpha\beta}, 
\sigma_z^{\alpha\beta}\right)
a_{i\alpha}^\dag a_{i\beta},
\end{equation}
with $\sigma_x$, $\sigma_y$, and $\sigma_z$ being the Pauli matrixes.
The dispersion relation of electrons is given by
\begin{equation}
E({\bf k}) = \epsilon_a + \frac1{N} 
\sum_{i \ne j}
t_{ij} e^{i{\bf k} \cdot \left({\bf R}_i-{\bf R}_j\right)} ,
\end{equation}
with $N$ the number of unit cells.
The bandwidth of $E({\bf k})$ is $W=(3\mbox{--}4)$~eV for cuprates. 
The third term in ${\cal H}$ is the superexchange interaction between nearest neighbors $\left<ij\right>$.
According to field theory, it arises from the virtual exchange of a pair excitation of electrons between UHB and LHB; \cite{FJO-ferro} it is certain that the superexchange interaction is present in not only insulating but also metallic phases.
In  the Hubbard model, the well-known value of $J=-4t^2/U$, with $t$ the transfer integral between nearest neighbors and $U$ the on-site repulsion,  can only be reproduced if the bandwidths of UHB and LHB are ignored.
When actual non-zero bandwidths are considered, $|J|$ is much smaller than $4t^2/U$. 
The experimental $J$ can only be explained beyond the Hubbard model, for example, by the $d$-$p$ model.\cite{exchange-reduction} In this paper, Eq.~(\ref{EqExpJ}) is assumed for $J$.
It is convenient to define
\begin{equation}\label{EqSuperJ}
J_s ({\bf q}) =
-2 J \left[\cos(q_xa)+\cos(q_ya)\right] ,
\end{equation}
with $J=- \mbox{(0.10--0.15)~eV}$ and $a$ the lattice constant. 
In the last term in ${\cal H}$, $U_\infty$ is
an infinitely large on-site repulsion such as $U_\infty/W \rightarrow+\infty$, which
is introduced to exclude any double occupancy.
Weak inter-layer couplings or 
quasi-two dimensionality is implicitly assumed so that
SC $T_c$ can be non-zero.

It is definite that, when $n\simeq 1$, $T_c$ of $d\gamma$ wave is much higher than $T_c$ of other waves. \cite{highTc1,highTc2} 
Provided that only a $d\gamma$-wave SC order parameter exists, which is denoted by $\Delta_{d\gamma} ({\bf k})$, the single-particle 
Green function in the Nambu representation is rigorously given by
\begin{widetext} 
\begin{eqnarray}\label{EqGreenNambu}
{\cal G}_{ \sigma}(i\varepsilon_n,{\bf k}) &=& 
- \int_{0}^{1/k_B T} \hskip-3pt d\tau e^{i\varepsilon_n \tau} 
\left<T_\tau \left(\begin{array}{c}
a_{{\bf k}\sigma}(0) \vspace{0.1cm} \\ a_{-{\bf k}-\sigma}^\dag(0)
\end{array}\right)
\left(
a_{{\bf k}\sigma}^\dag(\tau) , \ a_{-{\bf k}-\sigma}(\tau)
\right) \right> 
\nonumber \\ &=&
\left(\begin{array}{cc}
i\varepsilon_n -E({\bf k}) + \mu - \Sigma_\sigma(i\varepsilon_n, {\bf k}) &
-\Delta_{d\gamma} ({\bf k}) \\
-\Delta_{d\gamma}^* ({\bf k}) & 
i\varepsilon_n + E({\bf k}) - \mu 
+ \Sigma_{-\sigma}(-i\varepsilon_n, {\bf k})
\end{array} \right) ^{-1} ,
\end{eqnarray}
\end{widetext}
with $a_{{\bf k}\sigma}^\dag(\tau)$ and $a_{{\bf k}\sigma}(\tau)$
being creation and annihilation operators in the Heisenberg representation,
$\mu$ the chemical potential, 
and $\Sigma_\sigma(i\varepsilon_n, {\bf k})$ the self-energy.
The determinant of
${\cal G}_{ \sigma}^{-1}(i\varepsilon_n,{\bf k})$
is given by
\begin{eqnarray}\label{EqGreen00}
\bigl|{\cal G}_{ \sigma}^{-1}(i\varepsilon_n,{\bf k})\bigr|
&=&
%= 
\bigl[ i\varepsilon_n -E({\bf k}) + \mu
- \Sigma_\sigma(i\varepsilon_n, {\bf k})\bigr]
\nonumber \\ &&  \times
\bigl[i\varepsilon_n + E({\bf k}) - \mu 
+ \Sigma_{-\sigma}(-i\varepsilon_n, {\bf k}) \bigr] 
\nonumber \\ && \quad
- \bigl|\Delta_{d\gamma} ({\bf k})\bigr|^2.
\end{eqnarray}
It follows that
\begin{equation}\label{EqGreen11}
{\cal G}_{ \sigma}(i\varepsilon_n,{\bf k})=
\left(\begin{array}{cc}
G_{\sigma}(i\varepsilon_n, {\bf k}) & 
F(i\varepsilon_n,{\bf k}) \\
F^*(i\varepsilon_n,{\bf k})& 
- G_{-\sigma}(-i\varepsilon_n, {\bf k})
\end{array} \right) ,
\end{equation}
with
\begin{equation}\label{EqGreen12}
G_{\sigma}(i\varepsilon_n, {\bf k})  =
\frac{i\varepsilon_n + E({\bf k}) - \mu 
+ \Sigma_{-\sigma}(-i\varepsilon_n, {\bf k})}
{\bigl|{\cal G}_{ \sigma}^{-1}(i\varepsilon_n,{\bf k})\bigr|},
\end{equation}
and
\begin{equation}\label{EqGreen22}
F(i\varepsilon_n,{\bf k})  =
\frac{\Delta_{d\gamma} ({\bf k})}
{\bigl|{\cal G}_{ \sigma}^{-1}(i\varepsilon_n,{\bf k})\bigr|}.
\end{equation}
%

%\subsection{Single-site self-energy}
\subsection{Mapping to the Anderson model}

The Feynman diagrams of the $t$-$J$ model are classified
into single-site and multi-site ones.
If only $U_\infty$ and the single-site Green function,
which is defined by
\begin{equation}\label{EqLocalR}
R_{\sigma}(i\varepsilon_n) = \frac1{N} \sum_{\bf k} 
G_\sigma(i\varepsilon_n, {\bf k}),
\end{equation}
appear in a diagram, it is a single-site one.
If an inter-site part of $G_\sigma(i\varepsilon_n, {\bf k})$,
$F(i\varepsilon_n, {\bf k})$, or $J$ appears in a diagram, 
it is a multi-site one. 
According to this classification, 
the self-energy is divided into single-site and multi-site ones:
\begin{equation}
\Sigma_\sigma(i\varepsilon_n, {\bf k})=
\tilde{\Sigma}_\sigma(i\varepsilon_n)
+\Delta \Sigma_\sigma(i\varepsilon_n, {\bf k}).
\end{equation}
The single-site self-energy $\tilde{\Sigma}_\sigma(i\varepsilon_n)$ 
is given by that of the Anderson model:
\begin{eqnarray}\label{EqAM}
\tilde{\cal H} \hskip-3pt &=& \hskip-3pt
\epsilon_d \sum_{i\sigma}  n_{d\sigma}
+ \sum_{{\bf k}\sigma} E_{c} ({\bf k}) c_{{\bf k}\sigma}^\dag c_{{\bf k}\sigma}
+ U_\infty n_{d\uparrow} n_{d\downarrow}
\nonumber \\ && 
+ \frac1{\sqrt{ \tilde{N} }} \sum_{{\bf k}\sigma} 
\left[ V_{\bf k} c_{{\bf k}\sigma}^\dag d_\sigma
+ V_{\bf k}^* d_\sigma^\dag c_{{\bf k}\sigma}\right] , \quad
\end{eqnarray}
with $n_{d\sigma}=d_{\sigma}^\dag d_{\sigma}$, $\epsilon_d$ the energy level of $d$ electrons, $E_c({\bf k})$ the dispersion relation of conduction electrons, $\tilde{N}$ the number of lattice sites, $V_{\bf k}$ the hybridization matrix between $d$ and conduction electrons, and the on-site $U_\infty$ being the same as that for the $t$-$J$ model.
The Anderson model is characterized by $\epsilon_d-\tilde{\mu}$, with
$\tilde{\mu}$ the chemical potential, and
the hybridization energy defined by
\begin{equation}\label{EqHybridization}
\Delta (\varepsilon) = 
\frac{\pi}{\tilde{N}}\sum_{\bf k}|V_{\bf k}|^2 
\delta[\varepsilon+\tilde{\mu} - E_c({\bf k})].
\end{equation}
When $\Delta (0)>0$, the FS exists in the conduction band $E_c({\bf k})$. 
Then, the ground state of the Anderson model is the FL at least when $\Delta (0)>0$.
Consider the Green function defined by
\begin{equation}\label{EqAndersonG}
\tilde{G}_\sigma (i\varepsilon_n) =
\frac1{\displaystyle
i\varepsilon_n + \tilde{\mu}-\epsilon_d 
- \tilde{\Sigma}_\sigma (i\varepsilon_n)
-\frac1{\pi} \hskip-2pt \int \hskip-3pt
d\varepsilon^\prime \hskip-1pt \frac{\Delta (\varepsilon^\prime) }
{i\varepsilon_n - \varepsilon^\prime} } .~
\end{equation}
Provided that $\epsilon_d - \tilde{\mu}$ and $\Delta(\varepsilon)$
are determined to satisfy
\begin{equation}\label{EqMap-G}
R_{\sigma}(i\varepsilon_n) =\tilde{G}_{\sigma}(i\varepsilon_n),
\end{equation}
the single-site $\tilde{\Sigma}_\sigma(i\varepsilon_n)$
is simply the self-energy of the Anderson model.
Equation~(\ref{EqMap-G}) is the mapping condition to the Anderson model;
the multi-site $\Delta \Sigma_\sigma(i\varepsilon_n, {\bf k})$ and $\Delta_{d\gamma} ({\bf k})$ should also be self-consistently calculated with the single-site $\tilde{\Sigma}_\sigma(i\varepsilon_n)$ to satisfy Eq.~(\ref{EqMap-G}). 

When $\Delta \Sigma_\sigma(i\varepsilon_n, {\bf k})$ and $\Delta_{d\gamma} ({\bf k})$ are ignored in the mapping condition (\ref{EqMap-G}), the approximation is simply the supreme SSA, the DMFT, or the DCPA, which considers all the single-site terms. In this paper, both $\Delta \Sigma_\sigma(i\varepsilon_n, {\bf k})$ and $\Delta_{d\gamma} ({\bf k})$ are considered to treat the SC state properly.

According to Eq.~(\ref{EqMap-G}), DOS of the $t$-$J$ model is the same as that of the Anderson model:
\begin{equation}\label{EqRhoSSA}
\rho (\varepsilon) 
= -\frac1{\pi}\mbox{Im} R_\sigma(\varepsilon+i0) 
= -\frac1{\pi}\mbox{Im} \tilde{G}_\sigma(\varepsilon+i0).
\end{equation}
The electron density of the $t$-$J$ model is also the same as that of the Anderson model:
$n=(1/N)\sum_{i\sigma}\left<n_{i\sigma}\right>=\sum_{\sigma}\left<n_{d\sigma}\right>$.
It follows from the mapping condition~(\ref{EqMap-G}) that
%\begin{subequations}\label{EqMap0}
\begin{equation}\label{EqMap1}
\epsilon_d - \tilde{\mu}=\epsilon_a - \mu ,
\end{equation}
and
\begin{equation}\label{EqMap2}
\Delta (\varepsilon) =
\mbox{Im} \tilde{\Sigma}_\sigma(\varepsilon \!+\! i0)
+ \frac{\pi\rho(\varepsilon)}
{\left[\mbox{Re}R_{\sigma}(\varepsilon \!+\! i0)\right]^2 \!+
\left[\pi\rho(\varepsilon) \right]^2}. 
\end{equation}
%\end{subequations}
%
When $T=0$~K, 
${\rm Im}\tilde{\Sigma}_\sigma (+i0) = 0$ so that
\begin{equation}\label{EqHybrid0}
\Delta(0) = \frac{\pi\rho(0)}
{\left[\mbox{Re}R_{\sigma}(+i0)\right]^2+
\left[\pi\rho(0) \right]^2} .
\end{equation}
When no SC gap opens, $\rho(0)>0$ so that $\Delta(0)>0$ and the ground state of the Anderson model is the FL.
When an SC gap opens, $\rho(0)=0$ so that $\Delta(0)=0$ or 
no FS exists.
When the SC gap is anisotropic, 
the ground state of the Anderson model is still the FL,
as is examined in Sec.~\ref{SecSG-AM}.
The Kondo temperature $T_{\rm K}$ or $k_BT_{\rm K}$ is a characteristic low-energy scale, which is precisely defined in Sec.~\ref{SecMutualInt}.
It is assumed in the following part of this paper that $T\ll T_{\rm K}$ and $T_c\ll T_{\rm K}$.
In the presence of an infinitesimally small Zeeman energy $-\sigma h$ in the Anderson model, the self-energy of the FL is expanded as
\begin{eqnarray}\label{EqExpandSSA}
\tilde{\Sigma}_\sigma(\varepsilon \!+\! i0) &=&
\tilde{\Sigma}_0 + \bigl(1-\tilde{\phi}_\gamma \bigr)\varepsilon
+ \bigl(1-\tilde{\phi}_s \bigr)\sigma h
\nonumber \\ && + O\bigl(\varepsilon^2/k_BT_{\rm K}\bigr)
+ O\bigl(k_BT^2/T_{\rm K}\bigr), \hskip25pt
\end{eqnarray}
with $\tilde{\Sigma}_0$, $\tilde{\phi}_\gamma$, and
$\tilde{\phi}_s$ all being real. When $n\simeq 1$, charge fluctuations are suppressed so that $\tilde{\phi}_\gamma\gg 1$, $\tilde{\phi}_s \gg 1$, and
\begin{equation}
\tilde{\phi}_s  /\tilde{\phi}_\gamma \simeq 2,
\end{equation}
i.e., the so called Wilson ratio is about 2.\cite{yosida-yamada}  

When $\Delta \Sigma_\sigma(i\varepsilon_n, {\bf k})$ is considered but $\Delta_{d\gamma} ({\bf k})$ is ignored in Eq.~(\ref{EqGreen11}), the Green function is given by
\begin{eqnarray}\label{EqGreen}
G_{\sigma}(i\varepsilon_n, {\bf k}) &=& 
\frac1{\tilde{\phi}_\gamma}\frac1{
i\varepsilon_n -\xi_0({\bf k})-
\bar{\Delta}\Sigma_\sigma(i\varepsilon_n, {\bf k})} 
\nonumber \\ && \quad
+ \mbox{[incoherent term]} ,
\end{eqnarray}
with
\begin{equation}
\xi_0({\bf k}) = \frac1{\tilde{\phi}_\gamma}  \bigl[
\tilde{\Sigma}_0 + E({\bf k} )- \mu \bigr] ,
\end{equation}
and 
\begin{equation}
\bar{\Delta}\Sigma_\sigma(i\varepsilon_n, {\bf k}) =
\frac1{\tilde{\phi}_\gamma} \Delta\Sigma_\sigma(i\varepsilon_n, {\bf k}) .
\end{equation}
The first term in Eq.~(\ref{EqGreen})
is the coherent term, which describes the Gutzwiller band with band-width $W^*= O(k_BT_{\rm K})$ at the top of LHB.
The second term in Eq.~(\ref{EqGreen}) is the incoherent term, which
describes LHB; UHB is infinitely high in the $t$-$J$-$U_\infty$ model.

It is convenient to define
\begin{equation}
\bar{\Delta}_{d\gamma} ({\bf k}) =
\frac1{\tilde{\phi}_\gamma}  \Delta_{d\gamma}({\bf k}) ,
\end{equation}
and
\begin{eqnarray}\label{EqD}
\bar{D}_{ \sigma}(i\varepsilon_n,{\bf k}) &=&
\bigl[ i\varepsilon_n -\xi_0({\bf k})
- \bar{\Delta}\Sigma_\sigma(i\varepsilon_n, {\bf k})\bigr]
\nonumber \\ && \times
\left[ i\varepsilon_n + \xi_0({\bf k})
+ \bar{\Delta}\Sigma_{-\sigma}(-i\varepsilon_n, {\bf k})\right]
\nonumber \\ && \quad
- \left|\bar{\Delta}_{d\gamma}({\bf k})\right|^2 .
\end{eqnarray}
When both of $\bar{\Delta} \Sigma_\sigma(i\varepsilon_n, {\bf k})$ and $\bar{\Delta}_{d\gamma} ({\bf k})$ are considered in Eq.~(\ref{EqGreen11}), the Green function  is given by
\begin{equation}\label{EqEventualGreen}
{\cal G}_{ \sigma} (i\varepsilon_n,{\bf k})=
\frac1{\tilde{\phi}_\gamma} \left(\hspace{-3pt}
\begin{array}{cc}
\bar{G}_{\sigma}(i\varepsilon_n, {\bf k}) & 
\bar{F}(i\varepsilon_n,{\bf k}) \\
\bar{F}^*(i\varepsilon_n,{\bf k})& 
- \bar{G}_{-\sigma}(-i\varepsilon_n, {\bf k})
\end{array} \hspace{-3pt}\right) ,
\end{equation}
 with
\begin{equation}
\bar{G}_{\sigma}(i\varepsilon_n, {\bf k}) =
\frac{i\varepsilon_n - \xi_0({\bf k})
- \bar{\Delta} \Sigma_\sigma (i\varepsilon_n, {\bf k})}
{\bar{D}_{\sigma}(i\varepsilon_n,{\bf k})}, 
\end{equation}
and
\begin{equation}
\bar{F}(i\varepsilon_n,{\bf k}) =
\frac{\bar{\Delta}_{d\gamma} ({\bf k})}
{\bar{D}_{ \sigma}(i\varepsilon_n,{\bf k})} .
\end{equation}
The incoherent part is ignored here and in the following part. 

\subsection{Intersite exchange interactions}
\label{SecMutualInt}

The polarization function in spin channels is also divided into
the single-site and multi-site ones:
\begin{equation}
\pi_s(i\omega_l, {\bf q}) = \tilde{\pi}_s(i\omega_l) 
+\Delta\pi_s(i\omega_l, {\bf q}) .
\end{equation}
The spin susceptibility of the Anderson model is given by
\begin{equation}\label{EqSusAnd}
\tilde{\chi}_s(i \omega_l) = \frac{ 2\tilde{\pi}_s(i\omega_l) }{
 1-U_\infty \tilde{\pi}_s(i\omega_l) } ,
\end{equation}
and that of the $t$-$J$ model is given by
\begin{equation}\label{EqSusT-J}
\chi_s(i\omega_l, {\bf q})=\frac{2\pi_s(i\omega_l, {\bf q})}{
1-\left[\frac1{4}J_s({\bf q})+  U_\infty \right]\pi_s(i\omega_l, {\bf q})},
\end{equation}
with $J_s({\bf q})$ defined by Eq.~(\ref{EqSuperJ}).
It should be noted that
\begin{equation}\label{EqDelPi1}
\tilde{\pi}_s(i\omega_l) = 1/U_\infty + O\left[1/U_\infty^2\tilde{\chi}_s(i \omega_l) \right], 
\end{equation}
and
\begin{equation}\label{EqDelPi2}
\Delta\pi_s(i\omega_l, {\bf q}) = O\left[1/U_\infty^2\chi_s(i\omega_l, {\bf q})\right].
\end{equation}
A physical picture for Kondo lattices is that local spin fluctuations on different sites interact with each other by an intersite exchange interaction. Then, an intersite exchange interaction $I_s(i\omega_l,{\bm q})$ is defined by
\begin{equation}\label{EqDEfIs}
\chi_s(i\omega_l, {\bf q})=\frac{\tilde{\chi}_s(i \omega_l)}
{1- \frac1{4}I_s (i\omega_l, {\bf q})\tilde{\chi}_s(i \omega_l)}.
\end{equation}
It follows from Eqs.~(\ref{EqSusAnd}), (\ref{EqSusT-J}), (\ref{EqDelPi1}), (\ref{EqDelPi2}), and (\ref{EqDEfIs}) that
\begin{equation}\label{EqIs}
I_s (i\omega_l, {\bf q}) = J_s({\bf q})+
2 U_\infty^2\Delta\pi_s(i\omega_l, {\bf q}) ,
\end{equation}
in the limit of $U_\infty/W \rightarrow +\infty$.

The Kondo temperature is defined by 
\begin{equation}\label{EqDefTK}
k_BT_{\rm K} = \lim_{T\rightarrow0~{\rm K}} \bigl[1/\tilde{\chi}_s(0)\bigr].
\end{equation}
Since the mapped Anderson model is determined for a given $T$, it includes the given $T$ as a  parameter in addition to the temperature $T$ itself. In Eq.~(\ref{EqDefTK}), the limit of $T\rightarrow 0~{\rm K}$ stands for the limit for the temperature $T$; the parameter $T$ should be the given $T$. The Kondo temperature $T_{\rm K}$ depends on the parameter $T$.

The main term of $2 U_\infty^2\Delta\pi_s(i\omega_l, {\bf q}) $ in Eq.~(\ref{EqIs}) is an exchange interaction arising from the virtual exchange of a pair excitation of Bogoliubov's quasi-particles in the SC state. When the reducible and irreducible three-point single-site vertex functions in spin channels are denoted by $\tilde{\Lambda}_s(i\varepsilon_n+i\omega_l,i\varepsilon_n; i\omega_l)$ and $\tilde{\lambda}_s(i\varepsilon_n+i\omega_l,i\varepsilon_n; i\omega_l)$, respectively, it follows that
\begin{equation}
\tilde{\Lambda}_s(i\varepsilon_n+i\omega_l,i\varepsilon_n; i\omega_l) =
\frac{
\tilde{\lambda}_s(i\varepsilon_n+i\omega_l,i\varepsilon_n; i\omega_l) }
{1 -U_\infty \tilde{\pi}_s(i\omega_l)} , 
\end{equation}
according to the Ward relation. \cite{ward}  Since
%\begin{equation}
$\tilde{\phi}_s=\tilde{\Lambda}_s(0,0; 0)$,
%\end{equation}
%
it follows that
\begin{equation}\label{Eq3pointVertex}
\tilde{\lambda}_s(i\varepsilon_n+i\omega_l,i\varepsilon_n     ; i\omega_l)
=\frac{2 \tilde{\phi}_s}{U_\infty\tilde{\chi}_s(i\omega_l)} , 
\end{equation}
for $\varepsilon_n\rightarrow 0$ and $\omega_l\rightarrow 0$.
When Eq.~(\ref{Eq3pointVertex}) is approximately used, 
the exchange interaction is given by
\begin{equation}\label{EqJQ}
J_Q(i\omega_l,{\bm q}) = \frac{4(\tilde{\phi}_s/\tilde{\phi}_\gamma)^2}
{\tilde{\chi}_s^2(i\omega_l)}
\left[P(i\omega_l,{\bm q}) - \tilde{P}(i\omega_l) \right],
\end{equation}
with
\begin{widetext}
\begin{equation}
P(i\omega_l,{\bm q}) =
-k_BT \sum_{\varepsilon_n}\frac1{N}\sum_{{\bm k}\sigma}
\bigl[\bar{G}_\sigma(i\varepsilon_n+i\omega_l,{\bf k}+{\bf q} )
\bar{G}_\sigma(i\varepsilon_n,{\bf k})
+ \bar{F}(i\varepsilon_n+i\omega_l,{\bf k}+{\bf q} )
\bar{F}(i\varepsilon_n,{\bf k})\bigr] ,
\end{equation}
\end{widetext}
which is derived in the random-phase approximation (RPA) for pair excitations of Bogoliubov's quasi-particles.
In Eq.~(\ref{EqJQ}), the single-site term,
\begin{equation}
\tilde{P}(i\omega_l) =
-k_BT \sum_{\varepsilon_n}
\bar{R}_\sigma(i\varepsilon_n+i\omega_l)
\bar{R}_\sigma(i\varepsilon_n) ,
\end{equation}
with 
\begin{equation}
\bar{R}_\sigma(i\varepsilon_n) = \frac1{N}\sum_{\bf k}
\bar{G}_\sigma(i\varepsilon_n,{\bf k}),
\end{equation}
is subtracted. 
When the so called mode-mode coupling term among various types of fluctuations, which is denoted by $- 4 \Lambda (i\omega_l,{\bm q})$, is included,
it follows that
\begin{equation}
I_s(i\omega_l,{\bm q}) =
J_s({\bm q})
+J_Q(i\omega_l,{\bm q})
- 4 \Lambda (i\omega_l,{\bm q}).
\end{equation}
The mode-mode coupling term corresponds to that in the so called self-consistent renormalization (SCR) theory of spin fluctuations,\cite{kawabata,moriya} which is relevant in the weakly coupling regime.

When Eq.~(\ref{Eq3pointVertex}) is approximately used, 
the mutual interaction mediated by spin fluctuations is given by
\begin{equation}\label{EqChi-J}
\frac1{4} (U \tilde{\lambda}_s)^2
[\chi_s(i\omega_l,{\bm q}) - \tilde{\chi}_s(i\omega_l)]=
\frac1{4} \tilde{\phi}_s^2 I_s^*(i\omega_l,{\bm q}), 
\end{equation}
with $\tilde{\lambda}_s$ standing for 
$\tilde{\lambda}_s(0,0; 0)$ and  
\begin{equation}\label{EqIsStar}
I_s^*(i\omega_l,{\bm q})= \frac{ I_s(i\omega_l,{\bm q})}{
1 - \frac1{4} I_s(i\omega_l,{\bm q}) \tilde{\chi}_s(i\omega_l)}.
\end{equation}
In Eq.~(\ref{EqChi-J}), the single-site term is
subtracted and two $\tilde{\phi}_s$ appear as effective
three-point vertex functions. In the vicinity of the Mott \mbox{M-I} 
transition, the spin-fluctuation-mediated interaction is 
simply the exchange interaction $I_s^*(i\omega_l,{\bm q})$.

\subsection{Multi-site self-energy}

The multi-site self-energy $\bar{\Delta}\Sigma_\sigma(\varepsilon+i0, {\bf k})$ is perturbatively calculated in terms of $I_s(i\omega_l,{\bm q})$ or $I_s^*(i\omega_l,{\bm q})$.
First of all, the Fock-type term of the superexchange interaction $J_s({\bf q})$ should be considered to stabilize the {\it normal} or {\it unperturbed} state.\cite{FJO-Hartree,FJO-RVB}  When $J_s({\bf q})$ is treated in the mean-field approximation, three types of order parameters are possible such as superconductivity, antiferromagnetism, and bond order (BO). This implies that antiferromagnetism and BO compete with superconductivity. Not only SC fluctuations but also antiferromagnetic and BO fluctuations should be considered in calculating the multi-site self-energy $\bar{\Delta}\Sigma_\sigma(\varepsilon+i0, {\bf k})$.

When $T=0$~K or $T$ is so low that intersite thermal fluctuations can never be developed, $\bar{\Delta}\Sigma_\sigma(\varepsilon+i0, {\bf k})$ is expanded such that
\begin{equation}\label{EqMultiSelfExpand}
\bar{\Delta}\Sigma_\sigma(\varepsilon+i0, {\bf k}) =
\frac1{\tilde{\phi}_\gamma} \left[
\Delta \Sigma_0({\bf k}) - \Delta\phi_\gamma ({\bf k}) \varepsilon +\cdots
\right].
\end{equation}
The dispersion relation of quasi-particles is given by
\begin{equation}\label{EqXi}
\xi({\bf k}) = \bigl[E^*({\bf k})- \mu \bigr]/\phi_\gamma ({\bf k}),
\end{equation}
with
\begin{equation}
E^*({\bf k}) = \tilde{\Sigma}_0 
+ E({\bf k} ) +\Delta \Sigma_0({\bf k}),
\end{equation}
and
\begin{equation}
\phi_\gamma ({\bf k}) = \tilde{\phi}_\gamma + 
\Delta\phi_\gamma ({\bf k}) .
\end{equation}
According to the FL relation, the specific heat at $T_c < T \ll T_K$ is approximately given by
\begin{equation}
C(T) = \gamma T + \cdots,
\end{equation}
with
\begin{eqnarray}\label{EqSpecificG}
\gamma &=& 
\frac{2}{3}\pi^2 k_B^2
\frac1{N} \sum_{\bf k} \delta\bigl[\mu - E^*({\bf k}) \bigr]
\phi_\gamma ({\bf k})
\nonumber \\ &=&
\frac{2}{3}\pi^2 k_B^2\phi_\gamma \rho(0).
\end{eqnarray}
Here, 
\begin{equation}
\phi_\gamma = \frac{  \sum_{\bf k} 
\delta\bigl[\mu- E^*({\bf k}) \bigr]\phi_\gamma ({\bf k})}
{  \sum_{\bf k} 
\delta\bigl[\mu - E^*({\bf k}) \bigr]},
\end{equation}
is an average of $\phi_\gamma ({\bf k})$ over the FS and
\begin{equation}
\rho(0)=
\frac1{N} \sum_{\bf k} \delta\bigl[\mu - E^*({\bf k}) \bigr].
\end{equation}
When the development of intersite quantum fluctuations is as large as
that of local quantum spin fluctuations, it is possible that
\begin{equation}\label{EqWislon}
\phi_\gamma \simeq 2 \tilde{\phi}_\gamma \simeq \tilde{\phi}_s. 
\end{equation}
It is also possible that this relation holds at not only $T>T_c$ but also $T<T_c$.

When intersite thermal fluctuations are much developed,
the expansion (\ref{EqMultiSelfExpand}) is never relevant;
precise $T$ and $\varepsilon$ dependences of 
$\bar{\Delta}\Sigma_\sigma(i\varepsilon_n, {\bf k})$
should be considered.
Normal-state anomalies at $T>T_c$ such as those observed in under-doped cuprate superconductors can only be explained in terms of deviations from the expansion (\ref{EqMultiSelfExpand}). For example, a large pseudo-gap opens provided that $-\mbox{Im}\bar{\Delta}\Sigma_\sigma(\varepsilon+i0,{\bf k})$ has a large and sharp peak at $\varepsilon=0$. \cite{FJO-PsGap1,FJO-PsGap2}

\subsection{Gap equation}

An average of $I_s^*(\omega+i0,{\bm q})$ given by Eq.~(\ref{EqIsStar}) 
over a low-energy region $|\omega|\alt k_BT_{\rm K}$ is expanded as
\begin{equation}
\mbox{Re} \bigl<I_s^*(\omega+i0,{\bm q})\bigr>_{\omega} =
\sum_{lm} I_{(l,m)}^* e^{i(q_x l +q_y m )a},
\end{equation}
with $\left<\cdots\right>_{\omega}$ standing for the average and
$l$ and $m$ being integers.
When the  nearest-neighbor component, 
\begin{equation}
 I_{(\pm1,0)}^*=I_{(0,\pm1)}^* \equiv I_1^* , 
\end{equation}
is only considered as the attractive interaction, two types of singlet 
superconductivity is possible such as anisotropic $s$ and $d\gamma$ waves.
In this paper, $d\gamma$-wave superconductivity is considered since $T_c$ 
of $d\gamma$ wave is much higher than $T_c$ of $s$ wave.
When $\epsilon_G(T)$ is defined by
\begin{equation}\label{EqEqDeltaDG}
\bar{\Delta}_{d\gamma}({\bf k}) = \frac1{4}  
\eta_{d\gamma}({\bf q})\epsilon_G(T) ,
\end{equation}
with
\begin{equation}
\eta_{d\gamma}({\bf q}) = \cos(k_x a) - \cos(k_ya),
\end{equation}
being the form factor of $d\gamma$-wave Cooper pairs,
the gap equation is simply given by
\begin{equation}
1 = \frac{3}{4}|I_1^*| 
\left(\tilde{\phi}_s / \tilde{\phi}_\gamma \right)^2
\hskip-2pt \frac1{N}\sum_{\bf k} 
\frac{\eta_{d\gamma}^2({\bf k})}{\bar{D}_{ \sigma}(i\varepsilon_n,{\bf k}) } ,
\end{equation}
with $\bar{D}_{ \sigma}(i\varepsilon_n,{\bf k})$ defined by Eq.~(\ref{EqD}),
where Eq.~(\ref{EqEqDeltaDG}) should be used as
$\bar{\Delta}_{d\gamma}({\bf k})$.
When the gap equation is self-consistently solved,
the Green function is eventually given by Eq.~(\ref{EqEventualGreen}).

In the Kondo-lattice theory developed in this paper, in principle, 
single-site properties such as
$\xi_0({\bf k})$, $\tilde{\phi}_\gamma$, and $\tilde{\phi}_s$, 
the multi-site self-energy 
$\bar{\Delta}\Sigma_\sigma(i\varepsilon_n, {\bf k})$, 
the intersite exchange interaction $I_1^*$ between nearest neighbors,
and the gap function $\epsilon_G(T)$ should be self-consistently calculated with each other. 
The framework is in parallel to that of the conventional theory of superconductivity except that the {\it unperturbed} state or the single-site properties should be self-consistently calculated in the supreme SSA with the multi-site or intersite properties.

\section{Reduction of the Kondo temperature}
\label{SecReductionTK}
\subsection{Spontaneous magnetizations}
\label{SecSG-AM}

In general, $\mbox{Re}R_{\sigma}(+i0)\ne 0$.
According to Eq.~(\ref{EqMap1}), when the SC gap opens below $T_c$, $\Delta (\varepsilon)$ has also a gap structure; $\Delta(0)=0$ at $T=0$~K.
Although $\Delta(\varepsilon)$ should be self-consistently determined to satisfy the mapping condition (\ref{EqMap-G}) or (\ref{EqMap1}), the gap structure of $\Delta(\varepsilon)$ is phenomenologically treated.
Since $\Delta(\varepsilon) \propto |\varepsilon|$ for small $|\varepsilon|$ at $T=0$~K for $d\gamma$-wave superconductivity,
\begin{equation}\label{EqPhDelta}
\Delta(\varepsilon) = \left\{\begin{array}{cc}
\Delta_1, &- D \le \varepsilon  \le - \epsilon_0 \vspace{0.15cm}\\
\displaystyle \Delta_0 + 
\left(\Delta_1-\Delta_0\right) |\varepsilon|/|\epsilon_0|, 
& -\epsilon_0 \le \varepsilon \le 0 \vspace{0.2cm}\\
\displaystyle \Delta_0 + 
\left(\Delta_2-\Delta_0\right)|\varepsilon|/|\epsilon_0|, 
& 0 \le \varepsilon \le \epsilon_0 \vspace{0.15cm}\\
\Delta_2, &\epsilon_0 \le \varepsilon \le  D \vspace{0.15cm}\\
0 ,& |\varepsilon| >D 
\end{array}\right. ,
\end{equation}
is assumed; $\epsilon_0 \simeq (1/2)\epsilon_G(0)$ and $D\simeq W$. 
According to Eq.~(\ref{EqMap1}),
$\Delta_1 \simeq O(W/\pi)$ and $\Delta_1 \gg \Delta_2$,
since the chemical potential is at the top of LHB.
When the SC gap opens, $\Delta_0=0$ at $T=0$~K. However, 
$\Delta_0$ is treated as another parameter.
It is assumed that $\Delta_1 \ge \Delta_0 \ge 0$.

Since the on-site $U_\infty$ is infinitely large, no doubly occupied configuration of $d$ electrons appears in any eigen-state.
Then, the lowest singlet state of the Anderson model is expanded such that \cite{com1/N}
\begin{eqnarray}\label{EqPhi}
\Phi_{s} &=& \Bigl[ A_0
+ \sum_{{\bf k}\sigma} 
A_{d;{\bf k}\sigma} d_{\sigma}^\dag c_{{\bf k}\sigma}
\nonumber \\ && 
+\sum_{{\bf k}\sigma} \sum_{{\bf p}\sigma^\prime}
A_{{\bf k}\sigma;{\bf p}\sigma^\prime} c_{{\bf k}\sigma}^\dag
c_{{\bf p}\sigma^\prime} + \cdots
\Bigr] \left|0\right> , \quad
\end{eqnarray}
with $\left|0\right>$ being the Fermi vacuum for conduction electrons with no $d$ electron.   It satisfies
\begin{equation}
\bigl(\tilde{\cal H} 
-\tilde{\mu} {\cal N} + {\cal H}_{\rm ext} \bigr)\Phi_{s} = E_s \Phi_{s},
\end{equation}
with $\tilde{\cal H} $ defined by Eq.~(\ref{EqAM}), 
\begin{equation}
{\cal N} = \sum_{\sigma}n_{d\sigma} +
\sum_{{\bf k}\sigma}c_{{\bf k}\sigma}^\dag c_{{\bf k}\sigma},
\end{equation}
\begin{equation}
{\cal H}_{\rm ext} = - \sum_{\sigma}\left(
\Delta \mu +\sigma h\right)n_{d\sigma} ,
\end{equation}
being infinitesimally small external fields, and
$E_s$ the energy of the singlet. 
When only $A_0$, $A_{d;{\bf k}\sigma}$, and 
$A_{{\bf k}\sigma;{\bf p}\sigma^\prime}$ are considered, 
it follows that\cite{comPhDelta} 
\begin{equation}
E_s = -  \frac{2}{\pi} \hskip-2pt
\int_{-D}^{0} \hskip-5pt d\epsilon \Delta(\epsilon)
\frac{E_d- \Delta\mu - \epsilon - E_s  }{
\left(E_d- \Delta\mu - \epsilon -E_s  \right)^2 
- h^2 } ,
\end{equation}
where $E_d$ is the energy of the lowest doublet; 
\begin{equation}
E_d \simeq E_0 +\epsilon_d -\tilde{\mu}
-\frac{\Delta_2}{\pi} \ln \frac{D}{\Delta_1},
\end{equation}
with $E_0$ the energy of the Fermi vacuum.

When $\Delta_0>0$, it is obvious that $E_s<E_d$. Even if $\Delta_0=0$,  $E_s<E_d$. The ground state of the mapped Anderson model is a singlet within the phenomenological model (\ref{EqPhDelta}), i.e., when $\Delta(\varepsilon) \propto |\varepsilon|$ for small $|\varepsilon|$.

The $d$ electron density is given by
\begin{equation}
n = - \frac{\partial E_s}{\partial \Delta\mu }.
%$n = - \partial E_s/\partial \Delta\mu $.
\end{equation}
It follows that
\begin{equation}\label{EqSingleBinding}
\frac{n}{1-n} = 
\frac{2}{\pi}\int_{-D}^{0} 
d\epsilon  
\frac{\Delta (\epsilon)}
{(E_d - \epsilon - E_s)^2 },
\end{equation}
which gives $E_s$ as a function of $n$.
The magnetization of $d$ electrons is given by
\begin{equation}
m = - \frac{\partial E_s}{\partial h},
%$m = - \partial E_s/\partial h$
\end{equation}
so that the susceptibility is given by
\begin{eqnarray}
\tilde{\chi}_s(0) &=&
\lim_{h\rightarrow0} \frac{m}{h}
\nonumber \\ &=&
\frac{4(1-n)}{\pi}
 \int_{-D}^{0} d\epsilon 
\frac{\Delta (\epsilon)}
{(E_d - \epsilon  - E_s)^3}.\qquad 
\end{eqnarray}

First, consider the case where
no gap structure is developed in $\Delta(\varepsilon)$,
i.e., $\Delta_0=\Delta_1$ or $\epsilon_0=0$. 
It follows that
%\[\frac{n_d}{1-n_d} =\frac{2\Delta_1}{\pi}
%\frac1{E_d - E_s }, \]
\begin{equation}\label{EqEs0}
E_s = E_d - \frac{2 \Delta_1}{\pi}\frac{1-n}{n}  
< E_d ,
\end{equation}
and
%\[\tilde{\chi}_s(0) = \frac{4(1-n_d)\Delta_1}{\pi}
%\frac1{(E_d - E_s)^2 } ,\]
\begin{equation}\label{EqChiS0}
\tilde{\chi}_s(0) =  \frac{\pi}{2\Delta_1}\frac{n^2}{1-n}.
\end{equation}
Here, $D\rightarrow +\infty$ is assumed since $D \gg \left|E_d-E_s\right|$.

\begin{figure}
\centerline{
\includegraphics[width=8.0cm]{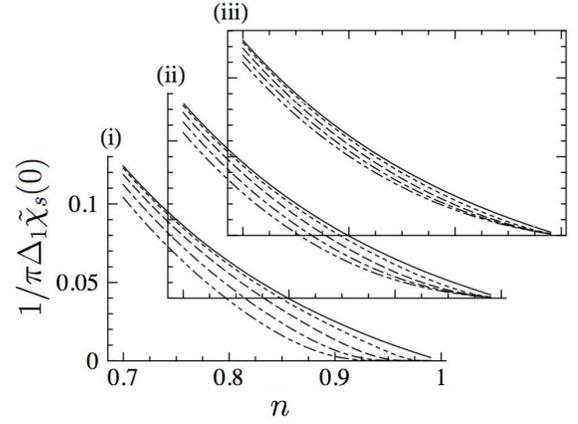}
}
\caption[1]{
$1/\pi \Delta_1 \tilde{\chi}_s(0)$ as a function of $n$;
(i) $\Delta_0/\Delta_1=0$, (ii) $\Delta_0/\Delta_1=0.1$, 
and (iii) $\Delta_0/\Delta_1=0.2$.
Solid, dotted, dashed, chain, and chain-dotted lines are for
$\varepsilon_0/\Delta_1=0$, 0.1, 0.2, 0.3, and 0.4, respectively;
every solid line shows Eq.~(\ref{EqChiS0}).
In any line, $1/\pi \Delta_1 \tilde{\chi}_s(0)>0$
even for $n \simeq 1$.
}
\label{fig_chis}
\end{figure}

When the gap structure is developed in $\Delta(\varepsilon)$ or
when $\Delta_0/\Delta_1$ is small and $\varepsilon_0/\Delta_1$ is large,
$\tilde{\chi}_s(0)$ is enhanced from Eq.~(\ref{EqChiS0}) or 
$1/\tilde{\chi}_s(0)$ is reduced. Figure~\ref{fig_chis} shows 
$1/\pi\Delta_1\tilde{\chi}_s(0)$ as a function of $n$
for various $\Delta_0/\Delta_1$ and $\varepsilon_0/\Delta_1$,
where $D\rightarrow +\infty$ is also assumed.

The N\'{e}el temperature $T_{\rm N}$ is defined by
\begin{equation}\label{EqAF-Instability}
\left[\frac1{\tilde{\chi}_s(0)} - \frac1{4}J_s({\bf q})
- \frac1{4} J_Q(0,{\bf q}) + \Lambda (0,{\bf q}) \right]_{T=T_{\rm N}}\!=0,
\end{equation}
where $T_{\rm N}$ should be maximized as a function of ${\bf q}$;
the ${\bf q}$ that gives the maximal $T_{\rm N}$ is the wave number ${\bf Q}$ of antiferromagnetic moments.
If Eq.~(\ref{EqAF-Instability}) gives a positive $T_{\rm N}$,
antiferromagnetism coexists with superconductivity.

According to Fig.~\ref{fig_chis}, $1/\tilde{\chi}_s(0)$ is largely reduced  when a large gap opens, for example, when $\Delta_0/\Delta_1 \simeq 0$, $\varepsilon_0/\Delta_1\agt 0.2$, and $n\agt0.9$.
When $1/\tilde{\chi}_s(0)$ is reduced, $J_Q(0,{\bf q})$ and $\Lambda (0,{\bf q})$ are also reduced. 
On the other hand, the superexchange interaction $J_s({\bf q})$ is scarcely reduced.
Provided that the gap is as large as $\epsilon_G(0)/\Delta_1 \agt 0.4$, therefore, it is probable that antiferromagnetism appears for $n\agt 0.9$, at least, at $T=0$~K or antiferromagnetism coexists with superconductivity for $n \agt0.9$.
When $J_s({\bf q})$ is given by Eq.~(\ref{EqSuperJ}),
the wave number of antiferromagnetic moments is ${\bf Q}\simeq (\pm\pi/a,\pm\pi/a)$.

\subsection{Small Knight shift}
\label{SecKnightShift}

The Knight shift is proportional to the static homogeneous susceptibility  $\chi_s(0,{\bf q}_0)$, with ${\bf q}_{0}$ standing for an infinitesimally small ${\bf q}$ such as $|{\bf q}|\rightarrow 0$. It follows that
\begin{equation}
\chi_s(0,{\bf q}_{0};T)  =
\frac1{\displaystyle 
X_0({\bf q}_{0})  + X_1({\bf q}_{0};T) },
\end{equation}
with
\begin{equation}
X_0({\bf q}_{0}) = - \frac1{4}J_s({\bf q}_{0})
+\frac{(\tilde{\phi}_s/\tilde{\phi}_\gamma)^2}
{\tilde{\chi}_s^2(0)}\tilde{P}(0)
+ \Lambda (0,{\bm q}_{0}),
\end{equation}
and
\begin{equation}
X_1({\bf q}_{0};T)  = \frac1{\tilde{\chi}_s(0)} 
- \frac{(\tilde{\phi}_s/\tilde{\phi}_\gamma)^2}
{\tilde{\chi}_s^2(0)}P(0,{\bm q}_{0}).
\end{equation}
The $T$-dependence of $X_1({\bf q}_{0};T)$ is only considered;
that of $X_0({\bf q}_{0};T)$ is ignored here.
When the {\bf k} dependence of $\phi_\gamma({\bf k})$ is ignored, or $\xi({\bf k})$ given by Eq.~(\ref{EqXi}) with $\phi_\gamma({\bf k})=\phi_\gamma$ is approximately used, it follows that 
\begin{equation}\label{EqYosida}
P(0, {\bf q}_{0}) =
\left(\frac{\tilde{\phi}_\gamma}{\phi_\gamma}\right)^2
\hskip-2pt
\frac1{N}\sum_{{\bm k}\sigma}
\frac1{k_BT}{\rm sech}^2\left[
\frac{\xi_{\rm sc}({\bf k})}{2k_BT}\right],
\end{equation}
with 
\begin{equation}
\xi_{\rm sc}({\bf k}) =
\sqrt{\xi^2({\bf k})
+(\tilde{\phi}_\gamma/\phi_\gamma)^2
\left|\mbox{$\frac1{2}$}\eta_\Gamma ({\bf k})\epsilon_G(T)\right|^2}.
\end{equation}
The function $P(0, {\bf q}_{0})$ is simply the Yosida function. 

When $T=0$~K, $|\epsilon_G(T)|>0$ so that $P(0,{\bf q}_{0})=0$ and
\begin{equation} 
X_1({\bf q}_{0}; 0\hskip1pt{\rm K}) = 
1/\tilde{\chi}_s(0; 0\hskip1pt{\rm K}) .
\end{equation}
Here, the $T$ dependence of $\tilde{\chi}_s(0)$ is explicitly shown. When $T>T_c$ but $T \alt T_{\rm K}$, on the other hand, $\epsilon_G(T)=0$ so that 
\begin{equation}
P(0,{\bf q}_{0}) \simeq 
\left(\frac{\tilde{\phi}_\gamma}{\phi_\gamma}\right)^2
\hskip-2pt
\frac1{N}\sum_{{\bf k}\sigma}\delta\bigl[\xi({\bf k})\bigr]
= \frac{2 \tilde{\phi}_\gamma^2}{\phi_\gamma}
\rho\bigl(0\bigr).
\end{equation}
When the hybridization energy $\Delta(\varepsilon)$ is constant, the local spin susceptibility at $T=0$~K is given by
\begin{equation}\label{EqSusFL}
\tilde{\chi}_s(0) = 2 \tilde{\phi}_s \rho(0) ,
\end{equation}
as is examined in Appendix~\ref{AppCompensation}. When $T\alt T_{\rm K}$ but no SC or no large pseudo-gap opens, i.e., $T\agt 2T_c$, Eq.~(\ref{EqSusFL}) approximately holds, so that
\begin{equation}
X_1({\bf q}_0; T) \simeq \bigl(1-
\tilde{\phi}_s/\phi_\gamma \bigr) 
/\tilde{\chi}_s(0; T).
\end{equation}
When Eq.~(\ref{EqWislon}) is satisfied, it follows that
\begin{equation}
X_1({\bf q}_0; 2T_c) \ll 1/\tilde{\chi}_s(0; 2T_c).
\end{equation}
If the reduction of $1/\tilde{\chi}_s(0;0\hskip1pt{\rm K})$ is large such that
\begin{equation}
1/\tilde{\chi}_s(0; 0\hskip1pt{\rm K})
= X_1({\bf q}_{0}; 0\hskip1pt{\rm K}) \ll 
1/\tilde{\chi}_s(0; 2T_c) ,
\end{equation}
the Knight shift is small; the $T$ dependence of $P(0,{\bf q}_{0})$ is cancelled by that of $1/\tilde{\chi}_s(0;T)$.
If the $T$ dependence of $1/\tilde{\chi}_s(0;T)$ is small below $T_c$ or $2T_c$, the Knight shift is large, as it is in conventional superconductors.

The Knight shift following the Yosida function is certainly evidence that the symmetry of superconductivity is singlet.
Since it can deviate from the Yosida function, however, a small Knight shift is never evidence that the symmetry of superconductivity is triplet if it occurs in the vicinity of the Mott \mbox{M-I} transition.

%\noindent 1  \\  2  \\  3  \\  4  %\\  5

\section{Phenomenological theory}
\label{SecSDW-lattice}

In the vicinity of the Mott transition, two types of antiferromagnetic orderings are possible: screw or helical spin structure and sinusoidal spin structure or spin density wave (SDW).\cite{Zachar,FJOmagStructureJ,FJOmagStructure}
When the nesting of the Fermi surface of quasi-particles is significant in a metal, SDW is stabilized,  which is presumably the case if antiferromagnetism occurs in metallic cuprates.
In general, SDW with wave number ${\bf Q}$ couples with charge density wave (CDW) with $2{\bf Q}$ and lattice distortion with $2{\bf Q}$.
Since the strong electron-phonon interaction arises from the modulation of the superexchange interaction.
\cite{FJO-ph1,FJO-ph2}  the coupling between SDW and lattice distortion must be strong but that between SDW and CDW is weak or is at least not strong. 
Then, Kondo-lattice theory should be extended to treat
the coexistence of superconductivity, SDW, and lattice distortion.
The extension is straightforward, as is discussed 
in Sec.~\ref{SecDiscussion}. For the sake of simplicity, however,
a phenomenological theory is developed in this section.

When an order parameter of SDW, that of lattice distortion, 
and a coupling constant between them are denoted by
\begin{equation}
m_{\bf Q} = m_{-{\bf Q}}^* = |m_{\bf Q}|e^{i\theta_m}, 
\end{equation}
\begin{equation}
u_{2{\bf Q}} = u_{-2{\bf Q}}^* = |u_{2{\bf Q}}|e^{i\theta_u},
\end{equation}
and
\begin{equation}
b_{2{\bf Q}} = b_{-2{\bf Q}}^* =|b_{2{\bf Q}}|e^{i\theta_b},
\end{equation}
with $\pm{\bf Q}$ and $\pm 2{\bf Q}$ being their ordering wave numbers, a coupling term in Landau's free energy is given by
\begin{eqnarray}\label{EqDeltaF}
\Delta F &=& 
b_{2{\bf Q}} m_{\bf Q}^2 u_{-2{\bf Q}}
+ b_{-2{\bf Q}} m_{-{\bf Q}}^2 u_{2{\bf Q}}
\nonumber \\ &=&
2|b_{2{\bf Q}} m_{\bf Q}^2 u_{2{\bf Q}}|
\cos(\theta_b+2\theta_m-\theta_u) ,\quad
\end{eqnarray}
The coupling term is minimal when 
\begin{equation}
\cos(\theta_b+2\theta_m-\theta_u)=-1.
\end{equation}
When an order parameter of superconductivity is denoted by $\Delta_{s}$, 
Landau's free energy to be considered is
\begin{eqnarray}\label{EqLandauFree1}
F(T) &=& 
F_0 + a_s (T)\left|\Delta_{s}\right|^2
+ a_m (T)\left|m_{\bf Q}\right|^2
\nonumber \\ &&
+ a_{2{\bf Q}} \left|u_{2{\bf Q}}\right|^2
- 2  |b_{2{\bf Q}} m_{\bf Q}^2 u_{2{\bf Q}}|
+ \frac1{2}c_s \left|\Delta_{s}\right|^4
\nonumber \\ &&
+ \frac1{2}c_m \left|m_{\bf Q}\right|^4 
+ c_{sm} \left|\Delta_{s}^2 m_{\bf Q}^2\right|+ \cdots ,
\quad
\end{eqnarray}
with
\begin{equation}
a_s(T) = \bar{a}_s k_B\left[T-T_c\right] ,
\end{equation}
and 
\begin{equation}
a_m(T) = \bar{a}_m k_B\left[T-T_m({\bf Q})\right] .
\end{equation}
Here, all the order parameters
are defined to be dimensionless. It is assumed 
that $a_{2{\bf Q}}>0$, $c_s>0$, $c_m >|b_{2{\bf Q}}|^2/a_{2{\bf Q}}$, 
$c_{sm}>0$, and $T_c > T_m({\bf Q})$.
According to the analysis in Sec.~\ref{SecSG-AM}, because of the reduction of $T_{\rm K}$ by the opening of an SC gap,
the coupling constant $c_{sm}$ can be much smaller 
in a superconductor in the vicinity of the Mott \mbox{M-I}
transition than it is in a conventional one. 
Since the free energy (\ref{EqLandauFree1}) is minimal when
\begin{equation}\label{Equ2Q}
\left|u_{2{\bf Q}}\right| = \frac{|b_{2{\bf Q}}|}{a_{2{\bf Q}}}
\left|m_{\bf Q}\right|^2 ,
\end{equation}
it follows that
\begin{eqnarray}
F (T) &=& 
F_0 + a_s (T)\left|\Delta_{s}\right|^2
+ a_m (T)\left|m_{\bf Q}\right|^2
+ \frac1{2}c_s \left|\Delta_{s}\right|^4
\nonumber \\ &&
+ \frac1{2}c_{\bf Q}\left|m_{\bf Q}\right|^4 
+ c_{sm} \left|\Delta_{s}^2 m_{\bf Q}^2\right| 
+ \cdots ,
\end{eqnarray}
with 
\begin{equation}
c_{{\bf Q}} = c_m - |b_{2{\bf Q}}|^2/a_{2{\bf Q}} >0.
\end{equation}
Below $T_c$, $|\Delta_{s}|$ is non-zero.
Define $T_{\rm N}^*({\bf Q})$ by
\begin{equation}
T_{\rm N}^*({\bf Q}) =
T_m ({\bf Q})- \frac{\bar{a}_{s}(c_{sm}/c_s)}
{\bar{a}_{m}-\bar{a}_{s}(c_{sm}/c_s)}
\left[T_c-T_m({\bf Q})\right] .
\end{equation}
When $T_m ({\bf Q})$ is large enough and
$c_{sm}$ is small enough, it follows that $T_{\rm N}^*({\bf Q})>0$.
In such a case, 
all of $|\Delta_{s}|$, $|m_{\bf Q}|$, and 
$|u_{2{\bf Q}}|$ are non-zero below $T_{\rm N}^*({\bf Q})$.

When $T_{\rm N}^*({\bf Q}) \le T \le T_c$, it follows that
\begin{equation}
|\Delta_{s}|^2 =
\frac{\bar{a}_s}{2c_s} k_B\left[T_c-T\right], 
\end{equation}
so that 
\begin{equation}
F(T)=F_0+F_1(T), 
\end{equation}
with
\begin{equation}
F_1(T) = -\frac{\bar{a}_s^2}{2c_s} 
k_B^2 \left[T_c-T\right]^2 .
\end{equation}
The specific heat jump at $T_c$ is given by
\begin{equation}
\Delta C_1 = \frac{\bar{a}_s^2}{c_s} k_B^2 T_c .
\end{equation}
When $0\le T \le T_{\rm N}^*({\bf Q})$, it follows that
\begin{equation}
|m_{\bf Q}|^2 = \frac{\bar{a}_{m}- \bar{a}_{s}(c_{sm}/c_s)}
{c_{\bf Q}-c_{sm}^2/c_s} k_B\left[T_m^*({\bf Q}) -T\right], 
\end{equation}
\begin{equation}
|\Delta_{s}|^2 =
\frac{\bar{a}_s }{2c_s} k_B\left[T_c-T\right]
- \frac{c_{sm}}{c_s} |m_{\bf Q}|^2 ,
\end{equation}
$|u_{2{\bf Q}}|$ is given by Eq.~(\ref{Equ2Q}), and
\begin{equation}
F(T)=F_0+F_1(T)+F_2(T;{\bf Q}), 
\end{equation}
with
\begin{equation}
F_2(T;{\bf Q}) = 
- \frac{\left[\bar{a}_{m}- \bar{a}_{s}(c_{sm}/c_s)\right]^2}
{2\left[c_{\bf Q}-c_{sm}^2/c_s\right]} 
k_B^2\left[T_m^*({\bf Q}) -T\right]^2. 
\end{equation}
The specific heat jump at $T_{\rm N}^*({\bf Q})$ is given by
\begin{equation}
\Delta C_2 = 
\frac{\left[\bar{a}_{m}- \bar{a}_{s}(c_{sm}/c_s)\right]^2}
{c_{\bf Q}-c_{sm}^2/c_s }k_B^2 T_m^*({\bf Q}). 
\end{equation}

When $T_m({\bf Q})$ is mainly determined by $J_s({\bf q})$ defined by Eq.~(\ref{EqSuperJ}), $T_m({\bf Q})$ is maximal at wave numbers in the vicinity of
\begin{equation}
{\bf Q}_M = \left(\pm \pi/a, \pm \pi/a \right).
\end{equation}
%  and $J_Q(i\omega_l,{\bm q})$ defined by Eq.~(\ref{EqJQ})
%for the square lattice. 
%Two spin-fluctuation modes with ${\bf q}_1$ and ${\bf q}_2$ couple with a phonon mode with ${\bf q}_1+{\bf q}_2$. 
When the enhancement of $T_m({\bf Q})$ by the strong electron-phonon interaction is large, $T_m({\bf Q})$ can be maximal
at wave numbers different from ${\bf Q}_M$.
Denote one of the wave numbers by ${\bf Q}_0$.
The N\'{e}el temperature is given by
\begin{equation}\label{EqSDW-TN}
T_{\rm N} = T_{\rm N}^*({\bf Q}_0) .
\end{equation}
The ordering wave number ${\bf Q}$ below $T_{\rm N}$ is determined by minimizing $F(T)$ or $F_2(T;{\bf Q})$ as a function of ${\bf Q}$.
When only the ${\bf Q}$ dependences of $T_m({\bf Q})$, $a_{2{\bf Q}}$, and $|b_{2{\bf Q}}|$ are considered, it is enough to minimize
\begin{equation}\label{EqF2A}
- \frac{ \left[T_m^*({\bf Q}) -T\right]^2}
{c_m - |b_{2{\bf Q}}|^2/a_{2{\bf Q}} -c_{sm}^2/c_s} ,
\end{equation}
instead of $F_2(T;{\bf Q})$.
When the electron-phonon interaction is strong, therefore, the ordering wave number of SDW can be different from ${\bf Q}_M$ at $T_{\rm N}$ and it can change with decreasing $T$ below $T_{\rm N}$.

\section{Checkerboard and Zero-temperature pseudo-gap in cuprate superconductors}
%\section{Application to cuprate superconductors}
\label{SecAnomalies}
The DOS depends on the position ${\bf r}$ in the checkerboard phase of cuprate superconductors, whose period is of about $4a\times 4a$. \cite{howald,hoffman1,hoffman2,fang,Hanaguri,ido4*4} 
The DOS defined by
\begin{equation}
\rho(\varepsilon; {\bf Q}_{\rm c}) =
\frac1{(2\pi)^2} \int d{\bf r}
e^{i{\bf Q}_{\rm c} \cdot {\bf r} }\rho(\varepsilon; {\bf r}),
\end{equation}
with ${\bf Q}_c\simeq(\pm\pi/2a,0)$ or $(0,\pm\pi/2a)$ being one of wave numbers of the checkerboard and $\rho(\varepsilon; {\bf r})$ being DOS  at ${\bf r}$ summed over spins, is almost symmetric with respect to $\varepsilon=0$ such that 
$\rho(\varepsilon; {\bf Q}_{\rm c}) \simeq \rho(-\varepsilon; {\bf Q}_{\rm c})$.\cite{howald}
It cannot be explained in terms of CDW by itself or CDW induced by lattice distortion, for which 
$\rho(\varepsilon; {\bf Q}_{\rm c})$ is almost asymmetric with respect to $\varepsilon=0$ such that 
$\rho(\varepsilon; {\bf Q}_{\rm c}) \simeq -\rho(-\varepsilon; {\bf Q}_{\rm c})$.
The ZTPG seems to be always accompanied by the checkerboard, \cite{Hanaguri,mcErloy1,mcErloy2} which implies that the ZTPG and the checkerboard below $T_c$ are different aspects of a phase.
Then, a possible interpretation for them is a double-{\bf Q} CDW or SDW in a conventional or exotic superconducting state: Cooper-pair density wave, a Wigner solid of Cooper pairs, and so on. \cite{Vojta,halperin,cdw,Chen2,Tesanovic,Davis,Anderson,Li,pereg-barnea,FJO-ZTPG}
As is studied in Secs.~\ref{SecSG-AM} and \ref{SecSDW-lattice},
SDW can coexist with $d\gamma$-wave superconductivity.
It is shown that the almost symmetric $\rho(\varepsilon; {\bf Q}_{\rm c})$ can be explained if CDW with an incommensurate ${\bf Q}_{\rm c}$ is induced by SDW with $\frac1{2}{\bf Q}_{\rm c}+{\bf G}$, ${\bf G}$ being a reciprocal lattice vector,  as a second-order effect of the SDW.\cite{FJO-ZTPG}
It is also shown that a similar fine structure to that of ZTPG can be reproduced if the coexistence of superconductivity and SDW is assumed.\cite{FJO-ZTPG}
The purpose of this section is to propose that, provided that $T<T_c$, the checkerboard and the ZTPG phases are different aspects of a coexistence phase of $d\gamma$-wave superconductivity, a double-{\bf Q} SDW, a double-{\bf Q} CDW, and a double-{\bf Q} lattice distortion.

Denote ordering wave numbers of SDW by
\begin{subequations}
\begin{equation}
{\bf Q} =  {\bf Q}_M 
+ {\bm \delta}(\pi/a), 
\end{equation}
with ${\bf Q}_M = \left(\pm \pi/a, \pm \pi/a \right)$ and
\begin{equation}
{\bm \delta} = \left(\pm \delta_x, \ \pm \delta_y\right) ,
\end{equation}
and those of the CDW and the lattice distortion by
\begin{equation}
2{\bf Q} = {\bm \delta} (2\pi/a) .
\end{equation}
\end{subequations}
According to previous papers, \cite{FJO-ph1,FJO-ph2} it is straightforward to show that $|b_{2{\bf Q}}|$ has a peculiar dependence on $2{\bf Q}$ or ${\bm \delta}$.
The mode with $2{\bf Q}={\bf Q}_M$ is the breathing mode.
Since it does not modulate the superexchange interaction, $|b_{2{\bf Q}}|$ are small around $2{\bf Q} = {\bf Q}_M$ or ${\bm \delta} = (\pm1/2,\pm1/2)$. 
The mode with $2{\bf Q}={\bf Q}_X$, with ${\bf Q}_X = \left(\pm \pi/a, 0 \right)$ or $\left(0, \pm \pi/a\right)$ is the half-breathing mode. 
Since it modulates strongly the superexchange interaction,
$|b_{2{\bf Q}}|$ is maximal around 
$2{\bf Q}={\bf Q}_X$, or
${\bm \delta}=\left(\pm 1/2, 0 \right)$ and  
${\bm \delta}=\left(0, \pm 1/2\right)$.
\cite{comQdepBq}
Then, it is probable that the enhanced $T_m({\bf Q})$ by the electron-phonon interaction is maximal on lines of
\begin{subequations}\label{EqLine0}
\begin{equation}\label{EqLine1}
0< |\delta_x|  <1/2, \quad \delta_y = 0,
\end{equation}
and 
\begin{equation}\label{EqLine2}
\delta_x=0, \quad 0< |\delta_y| < 1/2,
\end{equation}
\end{subequations}
in the plane of $(\delta_x,\delta_y)$ and the free energy (\ref{EqF2A}) is minimal on the lines.
It is possible that the free energy (\ref{EqF2A}) is minimal 
around midpoints of the lines such as
\begin{subequations}\label{EqAFM-Delta0}
\begin{equation}\label{EqAFM-Delta1}
{\bm \delta}_1 = \left( \pm 1/4, 0 \right),
\end{equation}
and
\begin{equation}\label{EqAFM-Delta2}
{\bm \delta}_2 = \left(0,  \pm 1/4\right).
\end{equation}
\end{subequations}
Two SDW's with ${\bf Q}_{1}={\bf Q}_M +{\bm \delta}_1(\pi/a)$ and ${\bf Q}_{2}={\bf Q}_M +{\bm \delta}_2(\pi/a)$
are equivalent to each other in the square lattice.
According to Kondo-lattice theory of SDW, \cite{FJOmagStructure} a double-${\bf Q}$ SDW with ${\bf Q}_{1}$ and ${\bf Q}_{2}$ must be stabilized rather than a single-${\bf Q}$ SDW with ${\bf Q}_{1}$ or ${\bf Q}_{2}$ since an additional stabilization energy appears when magnetization of different ${\bf Q}$ components are orthogonal to each other in the double-${\bf Q}$ SDW. 
The double-${\bf Q}$ SDW with ${\bf Q}_{1}$ and ${\bf Q}_{2}$ is inevitably accompanied by double-$2{\bf Q}$ lattice distortion and CDW with $2{\bf Q}_{1}$ and $2{\bf Q}_{2}$. 
The period of the double-$2{\bf Q}$ lattice distortion and CDW, which is of about $4a\times 4a$, is the observed period of the checkerboard structure

According to the previous paper, \cite{FJO-ph2}  phonons with $2{\bf Q}_{1}={\bm \delta}_1(2\pi/a)$ and $2{\bf Q}_{2}={\bm \delta}_2(2\pi/a)$, which are of the Cu-O bond-stretching mode, can become soft provided that antiferromagnetic spin fluctuations with ${\bf Q}_{1}$ and ${\bf Q}_{2}$ are developed. 
The softening of the bond-stretching mode is actually observed.\cite{pintschovius,reznik} 
In reverse, the softening enhances the instability of the double-${\bf Q}$ SDW.

If order parameters of $d\gamma$-wave superconductivity and the double-{\bf Q} SDW coexist homogeneously with each other, the dispersion relation of quasi-particles is cooperatively renormalized by the two order parameters or the opening of two types of gaps. Then, DOS can have a fine gap structure.\cite{FJO-ZTPG}
The fine structure must be ZTPG.

\section{Discussion}
\label{SecDiscussion}

It is straightforward to extend Kondo-lattice theory to study the coexistence of superconductivity, SDW, lattice distortion, and CDW.
When the wave number ${\bf Q}$ of SDW is incommensurate,
the Green function is given by
\begin{equation}\label{EqGreenSC-DW}
{\cal G}_{ \sigma}(i\varepsilon_n,{\bf k}) = 
- \!\int_{0}^{1/k_B T} \hspace*{-12pt} d\tau e^{i\varepsilon_n \tau} 
\left<T_\tau
\hat{A}_{{\bf k}\sigma}(0) ,
\hat{A}_{{\bf k}\sigma}^\dag(\tau) \right> ,
\end{equation}
with $\hat{A}_{{\bf k}\sigma}$ being 
the Hermitian conjugate of
\begin{eqnarray}\label{EqOpA}
\hat{A}_{{\bf k}\sigma}^\dag\hskip-3pt &=& \hskip-3pt
\bigl( \cdots,\hskip1pt
a_{{\bf k}-2{\bf Q}\hskip1pt \sigma}^\dag ,\hskip1pt
a_{{\bf k}-{\bf Q}\hskip1pt \sigma}^\dag ,\hskip1pt
a_{{\bf k}\hskip1pt \sigma}^\dag , \hskip1pt
a_{{\bf k}+{\bf Q}\hskip1pt \sigma}^\dag,\hskip1pt 
a_{{\bf k}+2{\bf Q}\hskip1pt \sigma}^\dag,
\nonumber \\ && \quad
\cdots\cdots, \hskip1pt
a_{-{\bf k}-2{\bf Q}\hskip1pt -\sigma},\hskip1pt
a_{-{\bf k}-{\bf Q}\hskip1pt -\sigma},\hskip1pt
a_{-{\bf k}\hskip1pt -\sigma},\hskip1pt
\nonumber \\ && \quad\quad
a_{-{\bf k}+{\bf Q}\hskip1pt -\sigma},\hskip1pt
a_{-{\bf k}+2{\bf Q}\hskip1pt -\sigma},\hskip1pt
\cdots \bigr) .
\end{eqnarray}
The self-energy and order parameters appear in the diagonal and off-diagonal parts, respectively, of Eq.~(\ref{EqGreenSC-DW}).
The formulation is almost in parallel with that in Sec.~\ref{SecKL-theory}.

Since the Fock-type term or the RVB mechanism, which suppresses the appearance of SDW, is reduced when the life-time width of quasi-particles is large, SDW more easily appears in more disordered systems than it does in less disordered systems. \cite{FJO-Hartree}
It is possible that disorder is much larger on surface CuO$_2$ layers than it is on deep  CuO$_2$ layers. In such a case, the checkerboard and ZTPG phase may exist only on disordered surface CuO$_2$ layers and it appears inhomogeneously on them. Experimentally, it is certain that the checkerboard and ZTPG phase appears on surface CuO$_2$ layers. It should be examined whether it can appear on deep CuO$_2$ layers.

The band center of LHB is around $\varepsilon\simeq -W/2$.
According to Hubbard's theory, \cite{Hubbard1,Hubbard2} $\rho(\varepsilon\simeq -W/2) \simeq 1/W$, so that $\rho(\varepsilon\simeq -W/2)\simeq 0.3~\mbox{eV}^{-1}$ for cuprates. 
When $\mbox{Im}\tilde{\Sigma}_\sigma(\varepsilon+i0)$ and $\mbox{Re}R_\sigma(\varepsilon+i0)$ are ignored in Eq.~(\ref{EqMap1}), $\Delta(\varepsilon\simeq -W/2) \simeq 1~\mbox{eV}$. 
When they are considered, it is plausible that
%
%\begin{equation}
$\Delta(\varepsilon\simeq -W/2) \simeq \frac1{3}\times 1~\mbox{eV}$,
%\end{equation}
i.e., $\Delta_1 \simeq 300~\mbox{meV}$.
On the other hand, $\Delta(0)$
given by Eq.~(\ref{EqHybrid0}) decreases as $\rho(0)$ decreases
provided that $\left|\mbox{Re}R_{\sigma}(+i0)\right| > \pi \rho(0)$.
When a pseudo-gap structure develops in $\rho(\varepsilon)$ above $T_c$,
$\Delta(\varepsilon)$ must also have a pseudo-gap structure.
Experimentally, a half of the pseudo-gap is as large as
$\varepsilon_0 \simeq 50~\mbox{meV}$.
Then, it is plausible that
\begin{equation}\label{EqLargeGap}
\varepsilon_0/\Delta_1 \simeq 0.2,
\end{equation}
in the pseudo-gap phase.
According to the results shown in Fig.~\ref{fig_chis},
it is possible that a double-{\bf Q} SDW or the checkerboard appear at 
$T \simeq T_c$ or $T\agt T_c$ if 
the pseudo-gap is so largely developed that not only Eq.~(\ref{EqLargeGap}) 
but also $\Delta_0/\Delta_1 \ll 1$ are satisfied.
The checkerboard is actually observed in the pseudo-gap phase above $T_c$.\cite{idoPG4*4}

Since two modes of antiferromagnetic spin-fluctuations with ${\bf q}_1$ and ${\bf q}_2$ inevitably couple with a phonon mode with ${\bf q}_1+{\bf q}_2$ due to the electron-phonon interaction, spin excitation spectra are renormalized by phonons.
It is interesting to examine the relevance of  such a spin-phonon
coupled mode to the so called resonance mode, which is observed in neutron inelastic scattering experiment. \cite{Rossat-Mignod}
It is also interesting to examine whether the Knight shift follows the Yosida function in under-doped cuprate superconductors, in which the pseudo-gap, the checkerboard, or ZTPG is observed.

When the Brillouin zone (BZ) is folded by SDW, CDW, and lattice distortion, Cooper pairs around one of the BZ edges or Cooper pairs with non-zero total momentum can coexist with Cooper pairs around the BZ center. \cite{FJO-ZTPG} 
A coexistence phase of such two types of Cooper pairs is
another type of  Cooper-pair density wave.
It is interesting to search for such a coexistence phase.

\section{Conclusion}
\label{SecConclusion}
According to Kondo-lattice theory of high-temperature superconductivity in cuprate superconductors, which is developed in a series of previous papers, the {\it normal} state is the Fermi liquid, which is mainly stabilized by the Kondo effect and the resonating valence bond (RVB) mechanism, and $d\gamma$-wave superconductivity occurs on CuO$_2$ planes mainly due to the superexchange interaction between-nearest neighbor Cu ions.  
In this paper, the Kondo-lattice theory is extended to treat properly the Kondo effect or the quenching effect of magnetic moments by single-site quantum spin fluctuations in a $d\gamma$-wave superconducting state. It is shown that, since the quenching effect is weakened by the opening of a $d\gamma$-wave superconducting gap, spin density wave (SDW) can appear in the superconducting state and the Knight shift can deviate from the Yosida function to be small. 

On the basis of the Kondo-lattice theory, together with the previous theory on the electron-phonon interaction on CuO$_2$ planes that arises from the modulation of the superexchange interaction by phonons, a phenomenological theory is developed  to study the coexistence of superconductivity, SDW, and lattice distortion. Because of the strong electron-phonon interaction,  the lattice distortion inevitably appears and the wave numbers of the SDW can be different from $(\pm\pi/a, \pm\pi/a)$, with $a$ being the lattice constant of CuO$_2$ planes, for which the superexchange interaction is maximal.

On the basis of the phenomenological theory,  
it is proposed that the checkerboard and the zero-temperature pseudo-gap (ZTPG) in the superconducting phase must be different aspects of a coexistence phase of $d\gamma$-wave superconductivity, a double-${\bf Q}$ SDW, a double-$2{\bf Q}$ lattice distortion, and a double-$2{\bf Q}$ charge density wave (CDW). 
The checkerboard can be explained if
the two wave numbers of the SDW are different from  $(\pm\pi/a, \pm \pi/a)$ because of the strong electron-phonon interaction, i.e., those of the SDW are ${\bf Q}_1 \simeq (\pm 3\pi/4a, \pm\pi/a)$ and ${\bf Q}_2 \simeq (\pm\pi/a, \pm 3\pi/4a)$ and those of the lattice distortion and the CDW are $2{\bf Q}_1 \simeq (\pm \pi/2a,0)$ and $2{\bf Q}_2 \simeq (0,\pm \pi/2a)$.  The fine structure of the ZTPG must be mainly due to the opening of superconducting and and SDW gaps.
On the basis of previous theories, it is discussed that
the stabilization mechanism of the double-{\bf Q} structure or the checkerboard rather than a single-{\bf Q} structure or the stripe is that an additional stabilization energy arises in the double-{\bf Q} SDW when magnetic moments of different {\bf Q} waves are orthogonal to each other,
the double-$2{\bf Q}$ CDW must be mainly induced by the double-${\bf Q}$ SDW rather than the double-$2{\bf Q}$ lattice distortion since the strong electron-phonon interaction, which arises from spin channels, can play no significant role in the induction of the CDW, and
the strong electron-phonon interaction can play no significant role either in the occurrence of $d\gamma$-wave superconductivity itself since the phonon-mediated attractive interaction is strong between next nearest neighbors but is weak between nearest neighbors.

\begin{acknowledgments} 
The author is thankful to M. Ido, M. Oda, and K. Nomura for useful discussions on experimental results.
\end{acknowledgments}

%\noindent 1 \\ 2  \\ 3  \\ 4 \\ 5 \\  6  %\\  7

\appendix
\section{Susceptibility of the Anderson model}
%\section{Anderson's compensation theorem}
\label{AppCompensation}

Consider the Anderson model (\ref{EqAM}) at $T=0$~K
in the presence of an infinitesimally small Zeeman energy $-\sigma h$.
The Green function for $d$ electrons is given by 
\begin{equation}\label{EqAM-G-App}
\tilde{G}_\sigma (i\varepsilon_n) =
\frac1{\displaystyle
i\varepsilon_n + \tilde{\mu}-\epsilon_d + \sigma h
- \tilde{\Sigma}_\sigma (i\varepsilon_n)
-L_\sigma(i\varepsilon_n) },
\end{equation}
and that for conduction electrons is given by
\begin{eqnarray}
G_{c\sigma}(i\varepsilon_n; {\bf k},{\bf k}^\prime) 
\hskip-3pt &=& \hskip-3pt
g_{c\sigma}(i\varepsilon_n,{\bf k})\delta_{{\bf k}{\bf k}^\prime}
+ g_{c\sigma}(i\varepsilon_n,{\bf k}) 
\nonumber \\ &&  \times
V_{\bf k} \tilde{G}_{\sigma}(i\varepsilon_n)V_{{\bf k}^\prime}^*
g_{c\sigma}(i\varepsilon_n,{\bf k}^\prime),\qquad
\end{eqnarray}
with
\begin{equation}
g_{c\sigma}(i\varepsilon_n,{\bf k}) =
\frac1{i\varepsilon_n +\tilde{\mu} -E_c({\bf k})} ,
\end{equation}
and 
\begin{equation}
L_\sigma(i\varepsilon_n) = \frac1{\tilde{N}} \sum_{\bf k}
|V_{\bf k}|^2g_{c\sigma}(i\varepsilon_n,{\bf k}) .
\end{equation}
The number of $d$ electrons is given by
\begin{equation}
n_{d\sigma} = 
\int_{-\infty}^{0} d\varepsilon 
\left(-\frac1{\pi}\right)\mbox{Im} \tilde{G}_\sigma(\varepsilon+i0), 
\end{equation}
and the change in the number of conduction electrons 
due to the hybridization is given by
\begin{equation}\label{EqD-nc}
\Delta n_{c\sigma} = 
\int_{-\infty}^{0} d\varepsilon 
\frac1{\pi} \mbox{Im}\left[
\frac{\partial L_\sigma(\varepsilon+i0) }{\partial \varepsilon}
\tilde{G}_{\sigma}(\varepsilon+i0) \right].
\end{equation}
The spin susceptibility due to $d$ electrons is defined by
\begin{equation}
\tilde{\chi}_s(0) = %\frac{d}{dh}
\bigl(d/dh\bigr)\sum_\sigma \sigma n_{d\sigma}, 
\end{equation}
and that due to conduction electrons is defined by
\begin{equation}
\Delta\tilde{\chi}_{s}(0) = %\frac{d}{dh}
\bigl(d/dh\bigr) \sum_\sigma 
\sigma \Delta n_{c\sigma}.
\end{equation}
It is straightforward to prove that \cite{shiba}
\begin{equation}
\tilde{\chi}_s(0) + \Delta\tilde{\chi}_{s}(0)
= 2\tilde{\phi}_s \rho(0),
\end{equation}
with $\tilde{\phi}_s$ the expansion coefficient defined 
by Eq.~(\ref{EqExpandSSA}) and $\rho(0)$ given by Eq.~(\ref{EqRhoSSA}) with Eq.~(\ref{EqMap-G}).

When the SC gap opens in the Hubbard or $t$-$J$ model,
$\rho(0)=0$ in the mapped Anderson model so that
\begin{equation}
\tilde{\chi}_s(0) + \Delta\tilde{\chi}_{s}(0)
= 0.
\end{equation}
On the other hand, $\tilde{\chi}_s(0)$ is enhanced, as is studied in Sec.~\ref{SecSG-AM}. Then, $\Delta\tilde{\chi}_{s}(0)$ is also enhanced in such a way that polarizations of conduction electrons exactly cancel those of $d$ electrons. 
Anderson's compensation theorem does not hold, i.e., $\Delta\tilde{\chi}_{s}(0)\ne 0$  in this case.

Consider a model where the hybridization energy defined by Eq.~(\ref{EqHybridization}) is given by
\begin{equation}
\Delta(\varepsilon) =  \Delta(0) \Gamma^2/(\varepsilon^2+\Gamma^2),
\end{equation}
or
%\begin{equation}
$L_\sigma(\varepsilon+i0) = 
\Delta(0) \Gamma/(\varepsilon+i\Gamma)$.
%\end{equation}
It follows that
\begin{equation}\label{EqDerivativeL}
\frac{\partial L_\sigma(\varepsilon+i0)}{\partial \varepsilon}
= -\Delta(0) \frac{\Gamma}{(\varepsilon+i\Gamma)^2} .
\end{equation}
In the limit of $\Gamma\rightarrow+\infty$, 
$\Delta(\varepsilon)$ is constant.
In such a case, Eq.~(\ref{EqDerivativeL}) vanishes so that $\Delta n_{c\sigma}=0$. Then, $\Delta\tilde{\chi}_{s}(0)=0$ or Anderson's compensation theorem holds.

%

%%%%%%%%%%%%%%%%%%%%%%%%%%%%%%%%%%%%
%%%%%%%%%%%%%%%%%%%%%%%%%%%%%%%%%%%%


\begin{thebibliography}{}
\bibitem{com1/N}
When the degeneracy of $d$ levels is denoted by $N_\lambda$, this expansion is simply the so called $1/N_\lambda$ expansion. The treatment is rigorous for a given $\Delta(\varepsilon)$ in the limit of $N_\lambda\rightarrow +\infty$.
\bibitem{comQdepBq}
Because of this peculiar wave-number dependence, the phonon-mediated attractive interaction is weak between nearest neighbors so that it can play no significant role in the occurrence of $d\gamma$-wave superconductivity itself.\cite{FJO-ph1,FJO-ph2}
\end{thebibliography}
\end{document}